\documentclass[aps,prb,onecolumn,amssymb,superscriptaddress,showpacs,preprint]{revtex4-1}
\usepackage{float}
\usepackage{xcolor}
\usepackage{graphicx}
\usepackage{dcolumn}
\usepackage{bm}
\usepackage{hyperref}
\usepackage{color}
\usepackage{amsmath}
\usepackage{amsfonts}

\begin{document}

\title{Stable Dirac semi-metal in the allotrope of \uppercase\expandafter{\romannumeral4} elements}

\author{Wendong Cao}
\affiliation{State Key Laboratory of Low-Dimensional Quantum Physics, Department of Physics, Tsinghua University, Beijing 100084, People's Republic of China}

\author{Peizhe Tang}
\email{peizhet@stanford.edu}
\affiliation{Department of Physics, McCullough Building, Stanford University, Stanford, California 94305-4045, USA}

\author{Shou-Cheng Zhang}
\affiliation{Department of Physics, McCullough Building, Stanford University, Stanford, California 94305-4045, USA}

\author{Wenhui Duan}
\email{dwh@phys.tsinghua.edu.cn}
\affiliation{State Key Laboratory of Low-Dimensional Quantum Physics, Department of Physics, Tsinghua University, Beijing 100084, People's Republic of China}
\affiliation{Institute for Advanced Study, Tsinghua University, Beijing 100084, People's Republic of China}
\affiliation{Collaborative Innovation Center of Quantum Matter, Beijing 100084, People's Republic of China}

\author{Angel Rubio}
\email{angel.rubio@mpsd.mpg.de}
\affiliation{Max Planck Institute for the Structure and Dynamics of Matter and Center for Free-Electron Laser Science, Luruper Chaussee 149, 22761 Hamburg, Germany}
\affiliation{Nano-Bio Spectroscopy group, Dpto.~F\'isica de Materiales, Universidad del Pa\'is Vasco, Centro de F\'isica de Materiales CSIC-UPV/EHU-MPC and DIPC, Av.~Tolosa 72, E-20018 San Sebasti\'an, Spain}

\date{\today}

\begin{abstract}
Three dimensional topological Dirac semi-metals represent a novel state of quantum matter with exotic electronic properties, in which a pair of Dirac points with the linear dispersion along all momentum directions exist in the bulk. Herein, by using the first principles calculations, we discover a new metastable allotrope of Ge and Sn in the staggered layered dumbbell structure, named as germancite and stancite, to be Dirac semi-metals with a pair of Dirac points on its rotation axis. On the surface parallel to the rotation axis, a pair of topologically non-trivial Fermi arcs are observed and a Lifshitz transition is found by tuning the Fermi level. Furthermore, the quantum thin film of germancite is found to be an intrinsic quantum spin Hall insulator. These discoveries suggest novel physical properties and future applications of the new metastable allotrope of Ge and Sn.
\end{abstract}


\maketitle

\section{Introduction}
The Dirac semi-metals, whose low energy physics can be described by three dimensional (3D) pseudorelativistic Dirac equation with the linear dispersion around the Fermi level \cite{Burkov2011}, have attracted lots of attention in recent days, owing to their exotic physical properties \cite{WangZJ-2012-Na3Bi,WangZJ-2013,li2010dynamical,potter2014quantum,Parameswaran2014} and large application potentials in the future \cite{Abrikosov1998,liang2014,He2014}. Current studies mainly focus on two types of Dirac semi-metals with both inversion symmetry and time-reversal (TR) symmetry. One is achieved at the critical point of a topological phase transition. This type of Dirac semi-metal is not protected by any topology and can be gapped easily via small perturbations \cite{sato2011-TlBiSSe,wu2013sudden,LiuJP2013}. In contrast, the other type is protected by the uniaxial rotation symmetry \cite{ChenF2012}, so is quite stable. And according to even or odd parity of the states at the axis of $C_n$ rotation, the symmetry protected Dirac semi-metals can be further classified as two subclasses \cite{YangBJ2014}. The first subclass has a single Dirac point (DP) at a time-reversal invariant momentum (TRIM) point on the rotation axis protected by the lattice symmetry \cite{YoungSM2012,Steinberg2014}, while the second one possesses non-trivial band inversion and has a pair of DPs on the rotation axis away from the TRIM points. For the materials of the second subclass (such as Na$_3$Bi \cite{WangZJ-2012-Na3Bi,liu2014discovery}, Cd$_3$As$_2$ \cite{WangZJ-2013,liu2014stable,Borisenko2014,yi2014evidence,liang2014,JeonSJ2014,He2014,Narayanan2015}, and some charge balanced compounds \cite{gibson20143d,du2014dirac}) the non-zero $\mathbb{Z}_{2}$ number can be well defined at the corresponding two dimensional (2D) plane of the Brillouin zone (BZ) \cite{Morimoto2014,Gorbar2015}. And due to the non-trivial topology, these stable Dirac semi-metals are regarded as a copy of Weyl semi-metals \cite{YangBJ2014}. Thus, its Fermi arcs are observed on the specific surfaces \cite{xu2015}, and a quantum oscillation of the topological property is expected to be achieved in the thin film with the change of thicknesses \cite{WangZJ-2013}.

In spite of these successful progresses, the 3D Dirac semi-metal materials either take uncommon lattice structures or contain heavy atoms, which are not compatible with current semiconductor industry. On the other hand, the group \uppercase\expandafter{\romannumeral4} elements, including C, Si, Ge, Sn and Pb, have been widely used in electronics and microelectronics. Generally, for some of the group \uppercase\expandafter{\romannumeral4} elements, the diamond structure is one of the most stable 3D forms at ambient conditions. However, under specific experimental growth conditions, various allotropes with exotic phyiscal and chemical properties are discovered experimentally. For example, the new orthorhombic allotrope of silicon, Si$_{24}$, is found to be a semiconductor with a direct gap of 1.3 eV at the $\Gamma$ point \cite{kim2015Si24}; and the 2D forms of silicene \cite{Seymur2009,Seymur2013-Sil,Seymure-sil-2014}, germanene \cite{Daviala2014,Chensi-2014} and stanene \cite{TangPz2014-stanene,Yong2013,zhu2015epitaxial} have been theoretically predicted to exist or experimentally grown on different substrates, which can be 2D topological insulators (TIs) and used as 2D field-effect transistors \cite{tao2015silicene}.

In this article, by using \emph{ab initio} density functional theory (DFT) with hybrid functional \cite{heyd2003hybrid}, we predict new 3D metastable allotropes for Ge and Sn with staggered layered dumbbell (SLD) structure, named as germancite and stancite; and discover that they are stable Dirac semi-metals with a pair of gapless DPs on the rotation axis of $C_3$ protected by the lattice symmetry. Similar to the conventional Dirac semi-metals, such as Na$_3$Bi and Cd$_3$As$_2$, the topologically non-trivial Fermi arcs can be observed on the surfaces parallel to the rotation axis in the germancite and stancite. And via tuning the Fermi level, we can observe a Lifshitz transition in the momentum space. More importantly for future applications, the thin film of the germancite is found to be an intrinsic 2D TI, and the ultrahigh mobility and giant magnetoresistance can be expected in these compounds due to the 3D linear dispersion.

\section{Methods}
The calculations were carried out by using DFT with the projector augmented wave method \cite{PhysRevB.50.17953,PhysRevB.59.1758}, as implemented in the Vienna \textit{ab initio} simulation package \cite{PhysRevB.54.11169}. Plane wave basis set with a kinetic energy cutoff of $\mathrm{250~eV}$ and $\mathrm{150~eV}$ was used for germancite and stancite respectively. The structure is allowed to fully relax until the residual forces are less than $1\times 10^{-3}~\mathrm{eV/\AA}$. The Monkhorst-Pack $k$ points are $9\times 9\times 9$. With the relaxed structure, the electronic calculation of germancite and stancite using hybrid functional HSE06 \cite{heyd2003hybrid} has been done with and without SOC. The maximally localized Wannier functions \cite{Mostofi2008685} are constructed to obtain the tight-binding Hamiltonian for the Green's function method \cite{0305-4608-15-4-009}, which is used to calculate the surface electronic spectrum and surface states.

\section{Results}
\begin{figure}
\centerline{ \includegraphics[clip,width=0.8\linewidth]{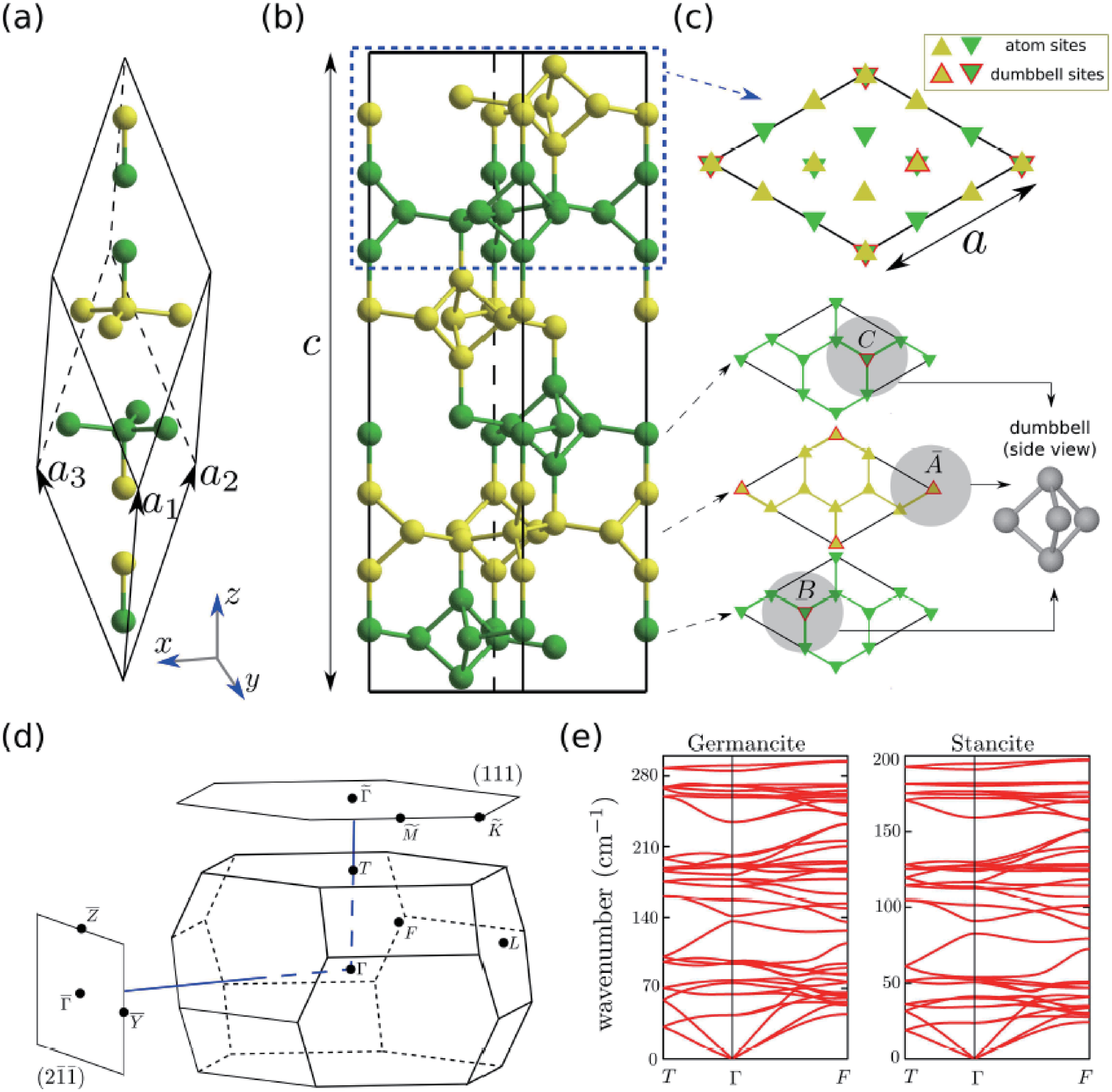}}
   \caption{(Color online) (a) The unit cell of the SLD structure with three private lattice vectors set as \textbf{a$_{1,2,3}$}. The balls in different colors stand for the same kind of atoms in different layers. (b) The side view and (c) top view of the SLD structure. The layers containing dumbbell (DB) structures are labelled. The letters ($A,B,C$) denote the positions of DB sites and the sign of bar is applied to distinguish between two trigonal lattices transformed to each other by inversion. As an example, the top view of two adjacent layers (marked by dashed blue lines) is shown. The DB structures are labeled by the grey shadow shown in the top view of a single layer, and the atoms in one DB structure are represented by grey balls. (d) The 3D Brillouin zone (BZ) of germancite and stancite.  The four inequivalent TRIM points are $\Gamma$ (0,0,0), $L$ (0,$\pi$,0), $F$ ($\pi$,$\pi$,0) and T ($\pi$,$\pi$,$\pi$). The hexagon and square, connected to $\Gamma$ by blue lines, show the 2D BZs projected to (111) and (2$\bar{1}\bar{1}$) surfaces respectively, and the high-symmetry $k$ points are labelled. (e) The phonon dispersion of germancite and stancite along high symmetry lines of 3D BZ.}
\label{fig:1}
\end{figure}

As shown in Fig. \ref{fig:1}, the germancite and stancite share the same rhombohedral crystal structure with the space group of $D_{3d}^6$ ($R\bar{3}c$) \cite{PhysRevB.90.085426}, which contains the spacial inversion symmetry and $C_3$ rotation symmetry along the trigonal axis (defined as $z$ axis). In one unit-cell, fourteen atoms bond with each others to form six atomic layers; and in each layer, one dumbbell site can be observed. To clearly visualize the SLD structure in the germancite and stancite, we plot the side view of the hexagonal lattice shown in Fig. \ref{fig:1}(b) and the top view from (111) direction in Fig. \ref{fig:1}(c). As the grey shadow shown, the layers containing dumbbell sites stack along (111) direction in the order of $\cdots B\bar{A}C\bar{B}A\bar{C}\cdots$. The interlayer interaction is the covalent bonding between adjacent layers, whose bond lengths are almost equal to those of intralayer bonding (the difference is about $0.03$\AA). Meanwhile, different from the diamond structure, the tetrahedral symmetry is absent in the SLD structure and the coupling here is not typical $sp^3$ hybridization. Furthermore, in order to test the structural stability, we calculate the phonon dispersion for the germancite and stancite shown in Fig. \ref{fig:1}(e). It can be seen that the frequencies of all modes are positive over the whole Brillouin zone, which indicates that the SLD structures are thermodynamically stable. Furthermore, compared with the other experimentally discovered metastable allotropes of Ge and Sn \cite{guloy2006guest,kiefer2010synthesis,PhysRevLett.110.085502,ja304380p,Ceylan2015407,PhysRevB.34.362}, the germancite and stancite share the same order of magnetite of the mass density and cohesive energies (see Supplemental Information for details), so we expect the germancite and stancite could be composed in the future experiments.

\begin{figure}
\centerline{ \includegraphics[clip,width=0.8\linewidth]{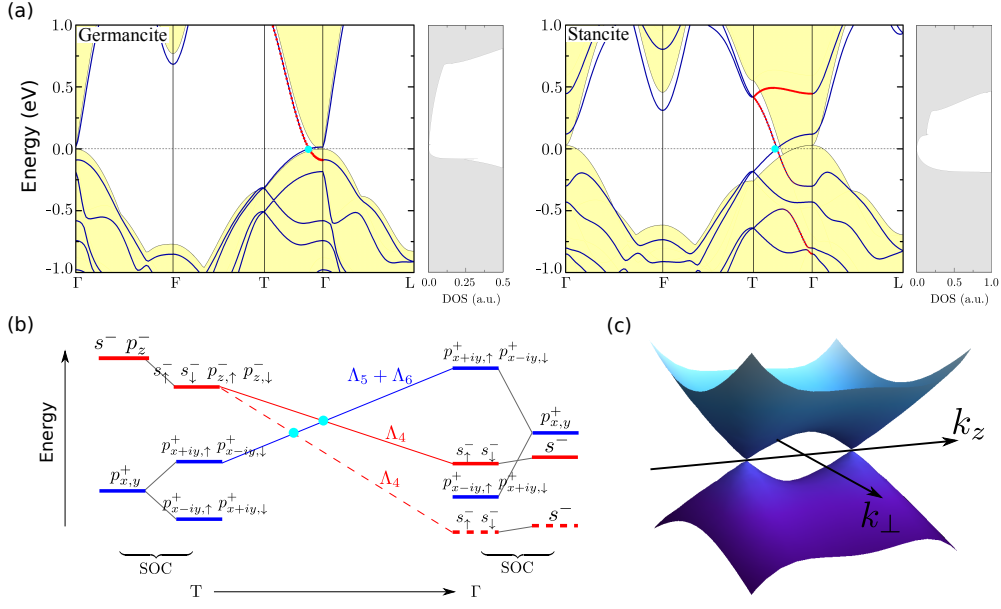}}
         \caption{(Color online) (a) The band structures of germancite (left) and stancite (right) along high symmetry lines with the corresponding DOS around the Fermi level (dashed horizontal line). In the k-path $\text{T}$-$\Gamma$, the size of the red dots represents the contribution from the atomic $s$ and $p_z$ orbitals. The cyan dots are the Dirac points at (0,0,$k_{z0}$), where $k_{z0}\approx 0.08 $ \AA$^{-1}$ and $\approx 0.18 $ \AA$^{-1}$ respectively. Shaded regions denote the calculated energy spectrum without SOC. (b) Schematic diagrams of the lowest conduction bands and highest valence bands from the $\text{T}$ point to the $\Gamma$ point for germancite and stancite. The black lines present the SOC effect at the $\text{T}$ and $\Gamma$ point. Between them, the red and blue lines denote doubly degenerate bands belonging to different irreducible representations, where the solid/dashed red line is for germancite/stancite. And the crossing points (solid cyan dots) correspond to those gapless Dirac points in (a) respectively. (c) Schematic band dispersion based on the effective $k\cdot p$ model for germancite and stancite. The $k_{\perp}$ direction refers to any axis perpendicular to the $k_{z}$ direction in the momentum space and the color becomes warmer, as the energy increases.}
\label{fig:2}
\end{figure}

The calculated electronic structures of the germancite and stancite around the Fermi level are shown in Fig. \ref{fig:2}(a), in which the solid lines and the yellow shadow stand for the bulk bands with and without spin-orbit coupling (SOC) respectively. It could be observed that: when the SOC effect is not included, the germancite is a conventional semi-metal whose bottom of the conduction bands and top of valence bands touch at the $\Gamma$ point with the parabolic dispersions; while for stancite, it is a metal whose band touching at the $\Gamma$ point is higher than the Fermi level. When the SOC effect is fully considered, our calculations indicate both germancite and stancite to be 3D Dirac semi-metals with a pair of DPs in the trigonal rotation axis (DP at (0,0,$\pm k_{z0}$)). Therefore, the low energy physics of this kind of materials can be described by the 3D Dirac-type Hamiltonian. And the schematic band structure based on the effective $k\cdot p$ model (see Supplemental Information for details) for germancite and stancite is shown in Fig. \ref{fig:2}(c), in which the pair of 3D DPs is clear.

To understand the physical origin of the 3D gapless Dirac Fermions in the SLD structure, we plot the schematic diagram of the band evolution for the germancte and stancite in Fig. \ref{fig:2}(b). In contrast to isotropic coupling in the diamond structure, the hybridizations in the layered SLD structure are anisotropic, in which the inter-layer couplings are relatively weaker than intra-layer couplings and the $p_z$ and $p_{x\pm iy}$ states are splited.  Furthermore, based on our calculations, the kind of anisotropic coupling will further shift down the anti-bonding state of $s$ orbital which is even lower than the bonding states of the $p_{x\pm iy}$ orbitals at the $\Gamma$ point. So the band inversion occurs at the $\Gamma$ point even without SOC effect, and the SOC herein just removes the degeneracy of $p_{x\pm iy}$ orbitals around the Fermi level. In the 2D BZ which contains the $\Gamma$ point and is perpendicular to the $\Gamma$-$\text{T}$ direction, the non zero $\mathbb{Z}_{2}$ topological number can be well defined. On the other hand, the $C_{3v}$ symmetry along the $\Gamma$-$\text{T}$ line contains one 2D ($\Lambda_{4}$) and two degenerate 1D ($\Lambda_{5}$, $\Lambda_{6}$) irreducible representations for its double space group \cite{koster1963properties}. As shown in the Fig. \ref{fig:2}(b), the two crossing bands at the Fermi level belong to $\Lambda_{5}+\Lambda_{6}$ and $\Lambda_{4}$ respectively.  So there is no coupling and a TR pair of 3D DPs can be observed at the Fermi level along the $\Gamma$-$\text{T}$ direction.

\begin{figure}
\centerline{ \includegraphics[clip,width=0.8\linewidth]{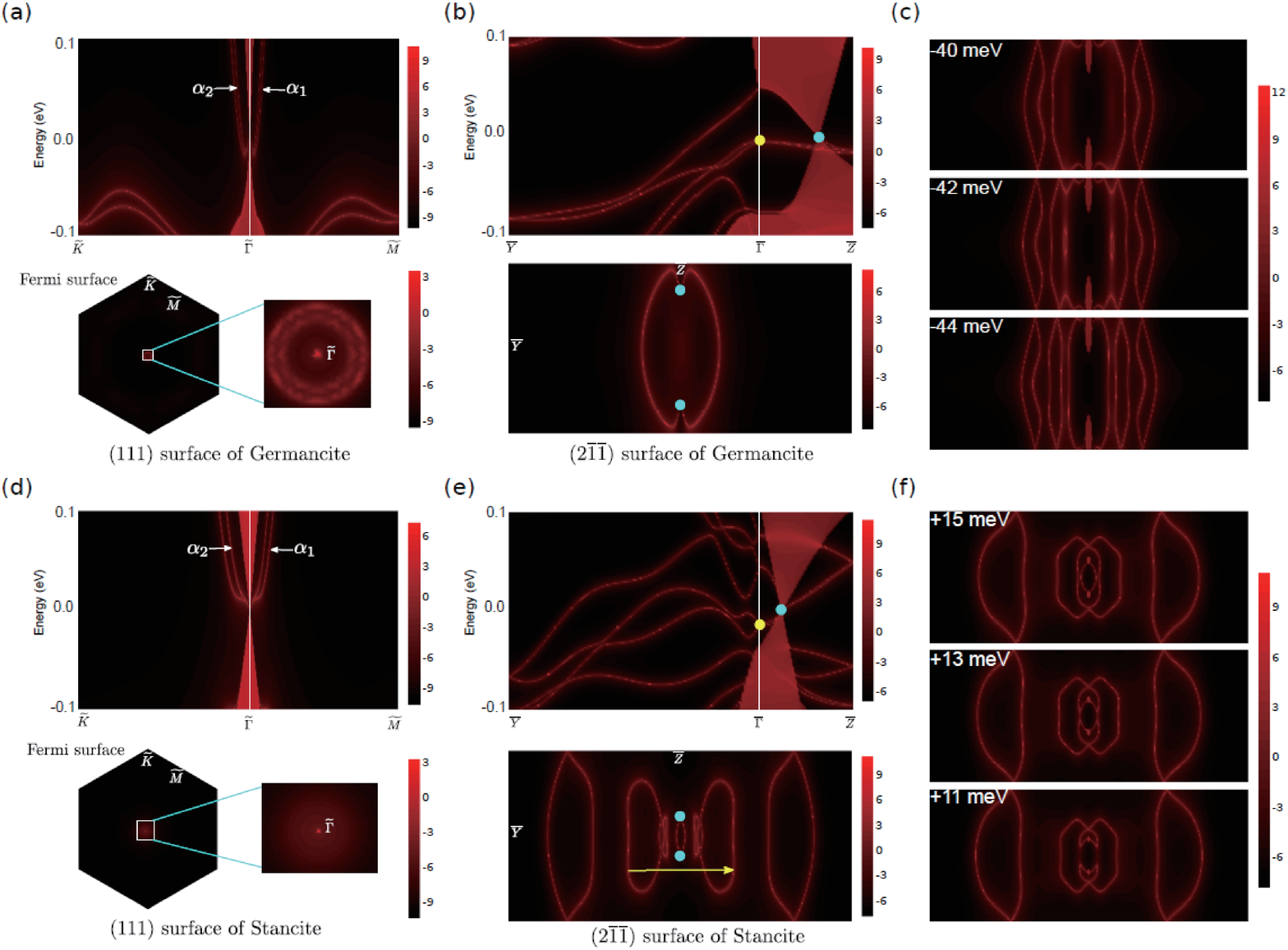}}
         \caption{(Color online) The electronic spectrum on the $(111)$ surface and its corresponding Fermi surface for (a) germanctie and (d) stancite respectively. Two bulk DPs are projected to the $\widetilde{\Gamma}$ point. The electronic spectrum on the $(2\bar{1}\bar{1})$ surface and its corresponding Fermi surface for (b) germanctie and (e) stancite respectively. The cyan dots label the projected DPs and the yellow dot represents the band crossing at the $\bar{\Gamma}$ point. On the Fermi surface, the Fermi arcs connect two projected DPs (cyan dots). For stancite $(2\bar{1}\bar{1})$ surface, the constant-energy contour is at $\epsilon_f-5.2$ meV, slightly away from the Fermi level, to distinguish the Fermi arcs. Stacking plots of constant-energy contours at different energies on its $(2\bar{1}\bar{1})$ surface of (c) germanctie and (f) stancite respectively. The Fermi level is set to be zero.}
\label{fig:3}
\end{figure}

Due to the non-trivial topology of 3D Dirac semi-metals, the projected 2D DPs and Fermi arcs are expected to be observed on some specific surfaces for the germancite and stancite. As shown in the Fig. \ref{fig:3}, by using the surface Green's function method \cite{0305-4608-15-4-009}, we study the electronic spectrum on the (111) and (2$\bar{1}\bar{1}$) surface whose BZs are perpendicular and parallel to the $\Gamma$-$\text{T}$ direction respectively. For the BZ of (111) surface, the pair of 3D DPs project to the $\widetilde{\Gamma}$ point as 2D Dirac cones (see Fig. \ref{fig:3}(a) and (d)); when the coupling between two projected 2D DPs is considered, a finite band gap could be easily obtained. Furthermore, besides the projected Dirac cones, we also observe the trivial surface states in the germancite and stancite ($\alpha_{1,2}$ states in the Fig. \ref{fig:3}(a) and (d)) which mainly originate from the dangling bonds on the (111) surface.

For the (2$\bar{1}\bar{1}$) surface of the germancite and stancite, the electronic structures are quite different. Because the BZ of (2$\bar{1}\bar{1}$) surface is parallel to the $\Gamma$-$\text{T}$ direction, the pair of 3D DPs are projected to different points (0,0,$\bar{\pm k_{z0}}$) which are marked by the cyan dots in the Fig. \ref{fig:3}(b) and (e). Between the projected DPs, a pair of the Fermi arcs could be observed clearly, which share the helical spin-texture and are not continuous at the projected points. This Fermi arcs originate from the non-trivial $\mathbb{Z}_{2}$ topology in the Dirac semi-metals. On any 2D plane in the bulk whose BZ is perpendicular to the $\Gamma$-$\text{T}$ direction with $-k_{z0}<k_z<k_{z0}$, the $\mathbb{Z}_{2}$ number is +1. Thus, in real space, the corresponding ``edge state" exist on the boundary. In the moment space, the BZ of the ``edge state'' corresponds to the line parallel to $\bar{Y}$-$\bar{\Gamma}$-$\bar{Y}$ with $-\bar{k_{z0}}$$<$$\bar{k_z}$$<$$\bar{k_{z0}}$, and its Fermi surface should be two points. After concluding all the contributions of planes with  $\mathbb{Z}_{2}$=1, the Fermi surface becomes a pair of the Fermi arcs on the BZ of (2$\bar{1}\bar{1}$) surface which connect the projected DPs. At the same time, on the (2$\bar{1}\bar{1}$) surface, the other surface states contributed by the dangling bond also exist. Via tuning the Fermi level, we could observe the hybridization between the non-trivial surface states and Fermi arcs (see Fig. \ref{fig:3}(c) and (f)), so a Lifshitz transition is found on the Fermi surface. Additionally, because the Fermi surface contours on the (2$\bar{1}\bar{1}$) surface contain roughly the same wave vector (see the yellow arrow in Fig. \ref{fig:3} (e)), the charge density wave or surface reconstruction is possible to be observed here. However, the surface coupling will not break the TR symmetry or change the bulk topology, the pair of Fermi arcs always exist.

\section{Discussion and Conclusion}
\begin{figure}
\centerline{ \includegraphics[clip,width=0.8\linewidth]{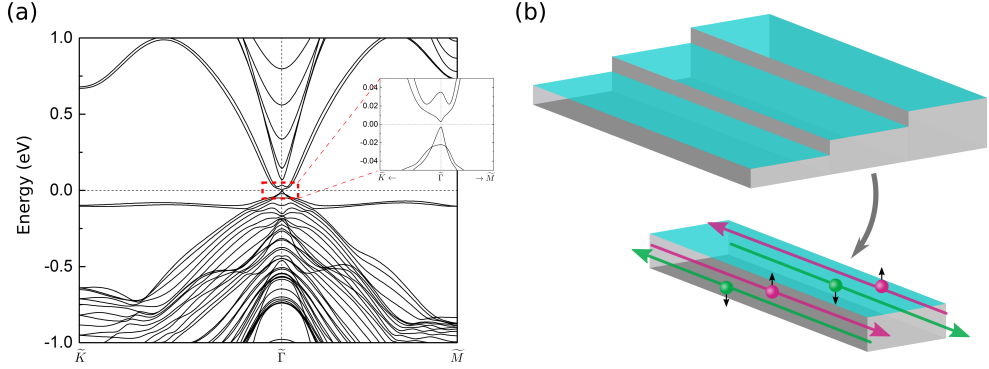}}
         \caption{(Color online) (a) Band structure of 16 layer germancite (111) film. The topologically nontrivial gap at the $\widetilde{\Gamma}$ can be seen in the inset. (b) Schematic device consisting of three germancite thin films with different thickness. The middle one is a QSH insulator, whereas the other two are topologically trivial. In the lower panel, the purple and green vectors stand for the spin-polarized current at the interfaces.}
\label{fig:4}
\end{figure}

Because of the compatibility with the traditional semiconductor devices and dissipationless edge transport, the realization of the quantum spin Hall (QSH) effect in the thin film of Ge attracts lots of attention recently. In a recent proposal \cite{Zhang2013-Ge-TI}, the non-trivial topology of the 2D thin film is induced by the large build-in electric field in the semiconductor interface, which may be difficult to control in real experiments. However, owing to the non-trivial topology of the Dirac semi-metal, the germancite (111) film may provide an opportunity for obtaining the QSH insulator. As discussed above, the topologically non-trivial band inversion occurs around the $\Gamma$ point in the germancite. So if we build 2D film with the proper thickness along the (111) direction, the band inversion may be restored at the $\widetilde{\Gamma}$ point and this thin film would become a QSH insulator. Figure \ref{fig:4} shows the electronic structure for germancite (111) film with the thickness of 14.5 nm (i.e., 16 layer). A small band gap (5.6 meV) opens at the $\widetilde{\Gamma}$ point due to the quantum confinement. To confirm our prediction, we calculate its $\mathbb{Z}_{2}$ number from the evolution of the Wannier charge centers (see Supplemental Information for details). It is found that the (111) thin film of the germancite is a 2D TI without applying the external electric field.

In conclusion, from DFT calculations with the hybrid functional, we predict the germancite and stancite with SLD structure are stable topological Dirac semi-metals protected by the rotation symmetry. And it is found that the Fermi arcs coexist with the trivial surface states on the surface plane parallel to the rotation axis of $C_3$, and a Lifshitz transition is observed when the Fermi level is tuned. Furthermore, we discover the (111) thin film of the germancite is a 2D TI without applying the external electric field which is important for future applications. Experimentally, the metastable allotropes of germanium has been synthesized through the oxidation of Ge$_9^{4-}$ Zintl anions in ionic liquids under ambient conditions \cite{guloy2006guest}. And owing to similar density and cohesive energy, we expect the germancite and stancite could be synthesized via the similar methods in the future.

\begin{acknowledgments}
We would like to thank S. Cahangirov and L. Xian for useful discussions. W.C. and W.D. acknowledge support from the Ministry of Science and Technology of China (Grant Nos. 2011CB606405 and 2011CB921901) and the National Natural Science Foundation of China (Grant No. 11334006). A.R. acknowledges financial support from the European Research Council Grant DYNamo (ERC-2010-AdG No. 267374) Spanish Grants (FIS2010-21282-C02-01), Grupos Consolidados UPV/EHU del Gobierno Vasco (IT578-13) and EC project CRONOS (280879-2 CRONOS CP-FP7). P.T. and S.-C.Z. acknowledge NSF under grant numbers DMR-1305677 and FAME, one of six centers of STARnet, a Semiconductor Research Corporation program sponsored by MARCO and DARPA. The calculations were done on the ``Explorer 100'' cluster system of Tsinghua University.
\end{acknowledgments}

W.C., P.T., W.D., S.-C.Z, and A.R. conceived and designed the project. W.C. and P.T. performed the $ab~initio$ calculations and theoretical analysis. P.T. and W.C. wrote the manuscript with the help from other authors. All authors contributed to the discussions. W.C. and P.T. contribute equally to this work.
%
\end{document}